# A Highly Adjustable Helical Beam: Design and Propagation Characteristic

Yuanhui Wen, Yujie Chen*, and Siyuan Yu

*State Key Laboratory of Optoelectronic Materials and Technologies, School of Electronics and Information Technology, Sun Yat-sen University, Guangzhou 510275, China*

*Corresponding author: chenyj69@mail.sysu.edu.cn*

Light fields with extraordinary propagation behaviours such as nondiffracting and self-bending are useful in optical delivery for energy, information, and even objects. A kind of helical beams is constructed here based on the caustic method. With appropriate design, the main lobe of these helical beams can be both well-confined and almost nondiffracting while moving along a helix with its radius, period, the number of rotations and main lobes highly adjustable. In addition, the main lobe contains almost half of the optical power and the peak intensity fluctuates below 15% during propagation. These promising characteristics may enable a variety of potential applications based on these beams.

Accelerating beams are light fields capable of self-bending in free space without any external potential. Such beams have gained great attention, not only for their counterintuitive propagation behaviour, but also for the versatile promising applications in the delivery of energy, information and even objects, or guiding using light, ranging from micro-machining[1], imaging[2,3], particle manipulation[4,5] to laser assisted guiding of particular processes like electric discharge[6] and plasma generation[7].

The first accelerating beam was demonstrated in 2007, in which a light field with initial Airy distribution is found to propagate along a parabolic trajectory[8,9]. Since then, considerable researches have been carried out to construct other kinds of accelerating beams propagating along various trajectories[10-13]. Among these, the helical trajectory is of particular interest for its remarkable rotating behaviour as well as the potential relationship with optical vortices[14-17]. The construction of helical beams at present mainly falls into two categories. One scheme is to superpose different order Bessel beams with the same angular velocity[18,19]. This does give a great freedom to construct helical beams with a variety of transverse profiles, but it is unable to ensure a satisfying main lobe like a well-confined hot spot to be constructed. Another scheme is to impose a spiral-shape phase or amplitude modulation[16,20-22], as well as the helico-conical phase[17,23,24] to the incident beam. However, although there is a clear main lobe of these beams propagating along helical trajectories, the characteristics of the main lobe are mostly lack of discussion, which is actually necessary and significant for practical applications.

In this work, a kind of helical beams is constructed based on the caustic method, whose main lobes move exactly along helical trajectories. The angular spectrum of this kind of beams is found to be conjugated with one of the helico-conical beams previously proposed[23], which suggests that the helical propagation behaviour of the helico-conical beam before the focus is easy to explain here. In order to improve the quality of the main lobe during the helical propagation, an additional ring-shaped amplitude modulation is introduced, which allows the main lobe to be well confined and almost nondiffracting at the same time. The helical beam is highly adjustable, including the radius and period of the helix, as well as the number of rotations and main lobes. In addition, other characterization of the main lobe is also given, in which the promising characteristics presented are attractive and desirable for considerable applications.

Note that the propagation of a scalar light field in linear isotropic medium can be described by an angular spectrum integral

$$E(X,Y,Z) = \int r(k_x, k_y) e^{i\varphi(k_x, k_y)} e^{i(k_x X + k_y Y + k_z Z)} dk_x dk_y \ , \quad (1)$$

where $E(X,Y,Z)$ is the field distribution at the space point $(X,Y,Z)$, while $r(k_x, k_y)$ and $\varphi(k_x, k_y)$ are the amplitude and phase of the angular spectrum in the initial plane with $k_x$ and $k_y$ to be the spatial frequency along $x-$ and $y-$ axes, respectively. Furthermore, under the paraxial approximation, the spatial frequency along the propagating direction can be expressed in the form of $k_z \approx k - (k_x^2 + k_y^2)/2k$ with $k$ to be the wave number in the linear medium, which is the case discussed in this paper and the wavelength of the light beam as well as the refractive index of the medium are set to be 632.8 nm and 1.5, respectively.

With appropriate design of the angular spectrum, a variety of light fields could be obtained based on Eq. (1), including the helical beams to be constructed, while the angular spectrum in the initial plane is required as the initial parameter. In order to determine the initial angular spectrum in need, we employ the caustic method to construct the helical beams, which associates the desired trajectory with an optical caustic, the envelope of a bundle of light rays[25]. In this case, the helical trajectory can be expressed mathematically by a parametric equation

$$X = u \cdot sin(t), \ Y = -u \cdot cos(t), \ Z = a \cdot t \ , \quad (2)$$

where $t$ is the parameter, $u$ the radius, and $2\pi a$ is the period, as shown in Fig. 1(a) with a red curve. The light ray in blue tangential to this trajectory can be described by its tangential equation

$$\frac{X - u \cdot sin(t)}{u \cdot cos(t)} = \frac{Y + u \cdot cos(t)}{u \cdot sin(t)} = \frac{Z - a \cdot t}{a} \ , \quad (3)$$

which intersects the initial plane at the point. Here the coordinate in the initial plane $(X, Y, 0)$ is denoted as $(x, y)$ for simplicity.

$$x = u \cdot [sin(t) - t \cdot cos(t)], \ y = -u \cdot [cos(t) + t \cdot sin(t)], \quad (4)$$

with the direction (spatial frequency) to be

$$\frac{k_x}{k} = \frac{\partial X}{\partial Z} = \frac{u \cdot cos(t)}{a} \ , \ \frac{k_y}{k} = \frac{\partial Y}{\partial Z} = \frac{u \cdot sin(t)}{a} \ , \quad (5)$$

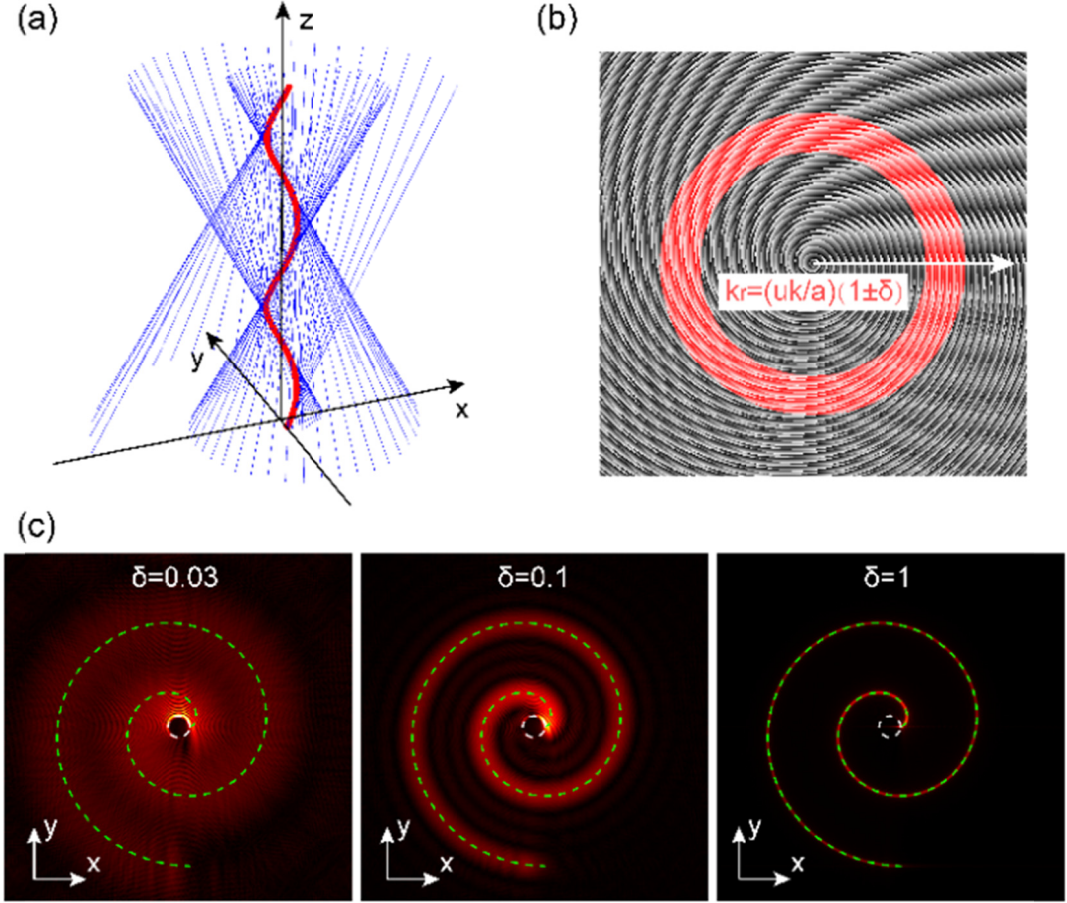

Fig. 1. Design of the helical beams. (a) Superposition of light rays in blue to form a helical caustic in red. (b) Angular spectrum of the helical beam in the initial plane, with the phase distribution in grayscale and an additional ring-shape amplitude distribution in red with a spatial frequency width described by $\delta$. (c) The initial field distribution corresponding to different spatial frequency width $\delta$.

or expressed in the polar coordinate as

$$k_r = \sqrt{k_x^2 + k_y^2} = \frac{u \cdot k}{a}, \quad k_\theta = arctan\left(\frac{k_y}{k_x}\right) = t \, . \quad (6)$$

By applying the stationary phase approximation to the Eq. (1) in the initial plane, we can obtain

$$\frac{\partial \varphi}{\partial k_x} = -x = -u \cdot [sin(k_\theta) - k_\theta \cdot cos(k_\theta)] \, , \quad (7)$$

$$\frac{\partial \varphi}{\partial k_y} = -y = u \cdot k_\theta \cdot [cos(k_\theta) + sin(k_\theta)] \, , \quad (8)$$

which actually can be explained as a mean approximation in the perspective of Wigner distribution[13,26]. Furthermore, by employing the chain rule between the polar coordinates and rectangular coordinates, the derivatives of phase function in polar coordinate system are

$$\frac{\partial \varphi}{\partial k_r} = \frac{\partial \varphi}{\partial k_x} \cdot \frac{\partial k_x}{\partial k_r} + \frac{\partial \varphi}{\partial k_y} \cdot \frac{\partial k_y}{\partial k_r} = u \cdot k_\theta \, , \quad (9)$$

$$\frac{\partial \varphi}{\partial k_\theta} = \frac{\partial \varphi}{\partial k_x} \cdot \frac{\partial k_x}{\partial k_\theta} + \frac{\partial \varphi}{\partial k_y} \cdot \frac{\partial k_y}{\partial k_\theta} = u \cdot k_r \, , \quad (10)$$

which results in an analytical form of the phase function as

$$\varphi(k_r, k_\theta) = u \cdot k_r \cdot k_\theta . \qquad (11)$$

This result is interesting because the radial and azimuthal dependence is inseparable, which reminds us of the previously proposed helico-conical beams[23] directly in the form of

$$\Psi(r, \theta) = l \cdot \theta \cdot (K - r/r_0) , \qquad (12)$$

In the case of $K = 0$, the phase function in Eq. (12) has almost the same form as Eq. (11), except for the minus, which leads to the helical propagation of the main lobe observed before the focus rather than after the focus in Ref. [23]. Therefore, a clear explanation is given here for the helico-conical beams in the case of $K = 0$ regarding the characteristics of spiral shapes [described by Eq. (4)] and helical propagation behaviour [described by Eq.(2)] based on our caustic analysis above.

The phase distribution of the angular spectrum described by Eq. (11) is presented in Fig. 1(b) with grayscale for a designed two-rotation helical beam. On the other hand, it is noted that the radial spatial frequency is constant according to Eq. (6), indicating that the amplitude distribution of the angular spectrum is a delta function and results in degeneration to a Bessel beam. To avoid this, a ring-shaped amplitude distribution with a certain width $\delta$ is introduced instead as shown in Fig. 1(b) in red, where $\delta$ is defined as the percentage of the half width over the central spatial frequency $u \cdot k/a$. Varying the spatial frequency width $\delta$, the corresponding field distributions in the initial plane are presented in Fig. 1(c). As can be seen, when $\delta$ is small, the initial field distribution tends to form a ring-like Bessel beam. As $\delta$ increases, the amplitude distribution evolves into a spiral shape with a clear main lobe, which is fitted precisely by Eq. (4) under the caustic analysis.

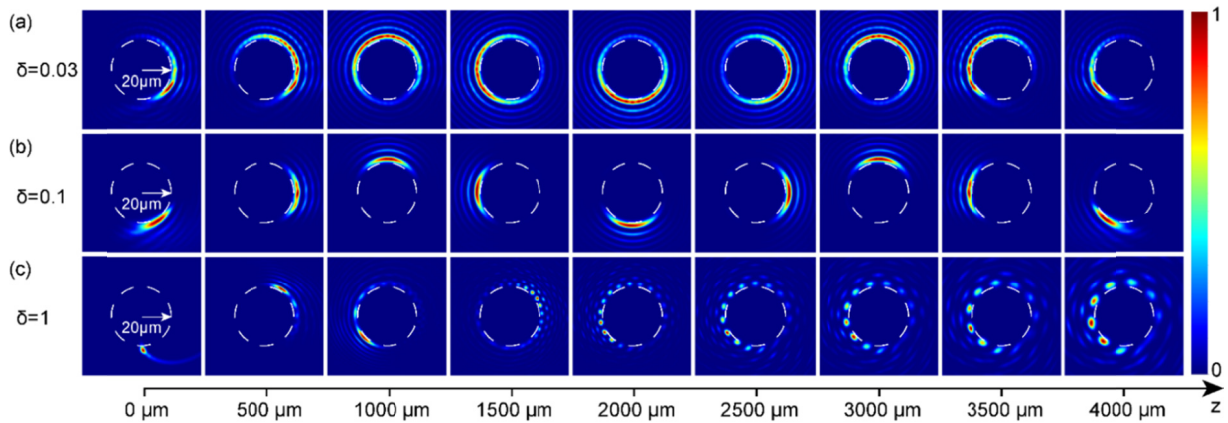

Fig.2. The cross-sectional intensity distribution of helical beams with spatial frequency width $\delta$ set to be (a) 0.03, (b) 0.1, and (c) 1, corresponding to Fig. 1(c). The radius and period of the predesigned helical trajectory are 20 and 2000 μm, respectively.

With the above angular spectrums in the initial plane, the propagation dynamics of these beams can be figured out by taking the numerical calculation of Eq. (1). The simulation results are shown in Fig. 2, corresponding to each initial field distribution shown in Fig. 1(c). Basically, the spatial frequency spectrum extending to a width $\delta$ has two effects on the beam. On one hand, the main lobe is better confined with a larger $\delta$, due to that more spatial frequency components are participated in the interference cancellation and thus leads to a better confinement in space. While on the other hand, as $\delta$ increases, the main lobe spreads out more quickly, because of that each radial spatial frequency actually corresponds to a different period of the helix according to Eq. (6). When the width $\delta$ is larger, the out-of-sync situation becomes worse. This is clearly shown in Fig. 2(c), in which the main lobe spreads out quickly and then the faster components catch up with the slower components, leading to the subsequent interference phenomenon. Therefore, selection of the width $\delta$ needs to balance both the confinement and the spreading of the main lobe. With an appropriate width $\delta$, a better helical propagation behaviour can be obtained as shown in Fig. 2(b), where the main lobe is well confined and almost propagates along a helical trajectory of 20 μm in radius and 2000 μm in period as designed. However, it is also noted that the main lobe deteriorates near the beginning and end of the predesigned caustic trajectory, owing to fewer spatial frequency components participate in the interference there.

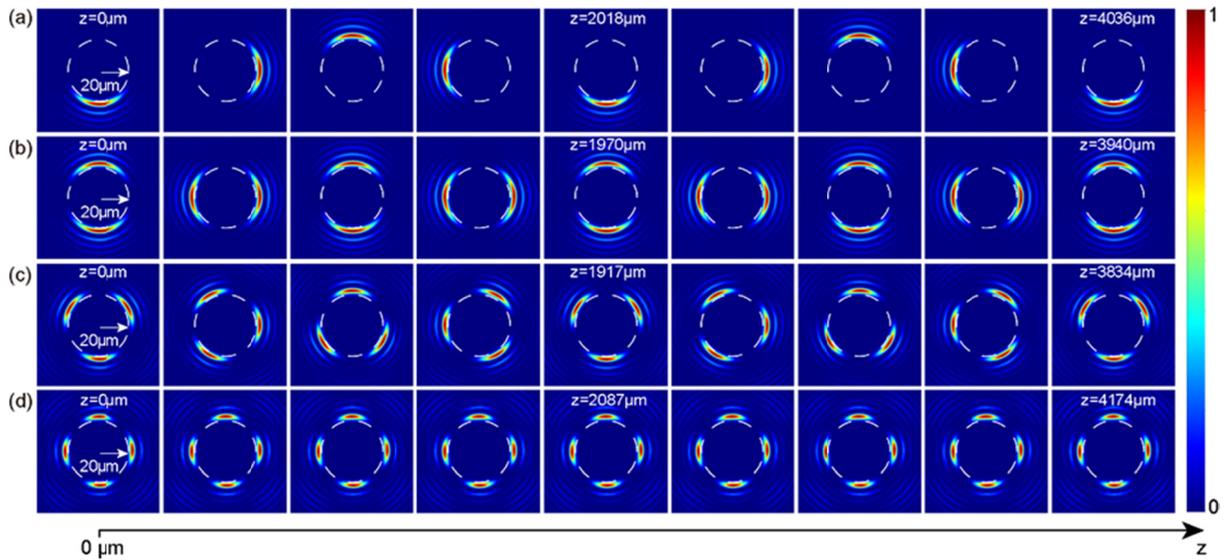

Fig.3. Propagation dynamics of helical beams with different number of main lobes. The helical beams presented includes (a) one-lobe, (b) two-lobe, (c) three-lobe, and (d) four-lobe, with slightly different period of 2018, 1970, 1917, and 2087 μm, respectively.

In order to improve the quality of the main lobe during the propagation along the desired trajectory, more spatial frequency components are included as all as those essential to the desired caustic. With this improvement, the main lobe becomes much better especially at the beginning and end of the required trajectory, as shown in Fig. 3(a). Moreover, the period of the helix is obtained clearly in this case, which is about 2018 μm, within the uncertain range of $(2000/1.1, 2000/0.9)$ μm due to the extension of the spatial frequency spectrum, and the period actually can be adjusted at will by simply altering the central spatial frequency. In addition to the radius, period and the number of rotations of the helix capable of tailoring freely in the previous caustic analysis, it is found that the number of main lobes is also adjustable. For demonstration, we have constructed helical beams with the number of main lobes varying from one to four, as shown in Fig. 3, based on the abovementioned helical beam. The propagation dynamics of these helical beams is attractive and desirable since all the main lobes are well confined and almost nondiffracting during propagation, which has potential applications for micro-manipulation, for instance, to trap different number of particles at the same time. Although the period of the helix is slightly different, it can be easily compensated by shifting the central spatial frequency as mentioned before.

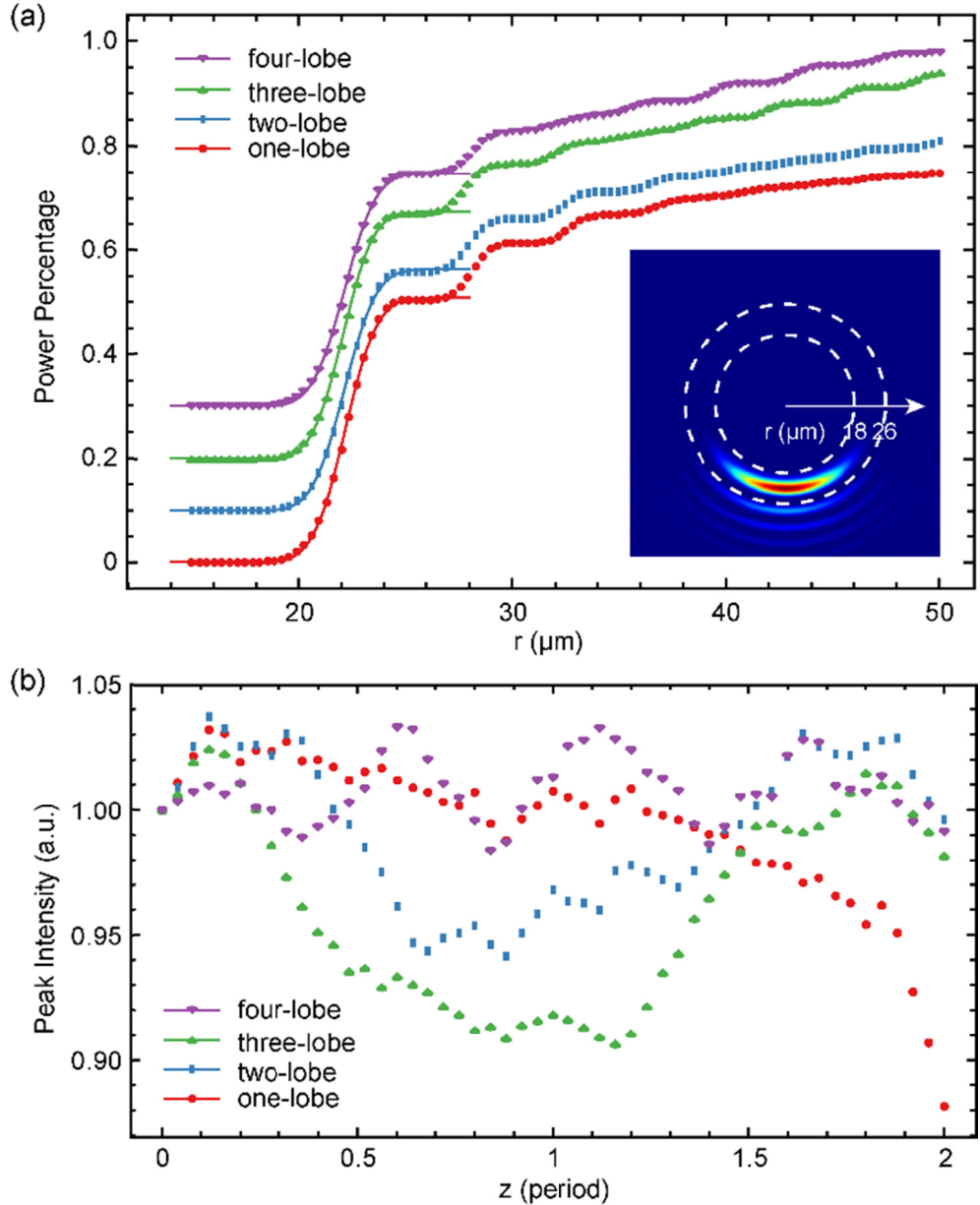

Fig. 4. Characterization of the helical beams with different number of main lobes in Fig. 3. (a) The percentage of power within a circle in radius of $r$, which is fitted by $0.254\{1 + \text{Erf}[0.6(x - 22.2)]\}$, $0.232\{1 + \text{Erf}[0.6(x - 22.1)]\}$, $0.236\{1 + \text{Erf}[0.63(x - 22.1)]\}$ and $0.224\{1 + \text{Erf}[0.6(x - 22.1)]\}$ for helical beams with one to four main lobes, respectively. Noted that the curves for two-lobe, three-lobe, and four-lobe helical beams are shifted upward by 0.1, 0.2, and 0.3 for a better visualization. The inset shows the first two flat states in the curve and contains the main lobe. (b) The fluctuation of the peak intensity during the two-rotation helical propagation, which is normalized by the initial peak intensity for a better comparison.

It is noted that the most important and common characterizations of such helical beams include the percentage of the power contained in the main lobe and the fluctuation of the peak intensity during propagation. Here the percentage of the power within a circle in radius of $r$ is calculated and presented in Fig. 4(a) for the constructed helical beams in Fig. 3. When $r$ is small, there is almost no optical power in the dark center. As $r$ increases and crosses the main lobe, the power rises quickly and then remains flat again after passing the main lobe. The variation is fitted well with an error function, which indicates a nice Gaussian intensity distribution of the main lobe along the radial direction, with a FWHM to be around 2.3 μm. From the fitted results, we can obtain that the percentage of the power contained in the main lobe for these helical beams with one to four lobes is 50.8%, 46.4%, 47.2%, and 44.8%, respectively. It is quite surprising that almost half of the optical power remain in the main lobe, which is almost as high as that of two-dimensional Airy beams. Moreover, the fluctuation of the peak intensity during propagation is within 15% (mostly within 10%) and might be further improved with an optimization.

In conclusion, we have constructed a kind of helical beams based on the caustic method with an additional ring-shape amplitude distribution. This kind of beams is related to one of the helico-conical beams and thus clearly explains its helical propagation behaviour before the focus. With appropriate design, the main lobe of these helical beams can be both well confined and almost nondiffracting while moving along a predesigned helical trajectory with its radius, period and the number of rotations highly adjustable. In addition, there are promising characteristics of these helical beams, including the number of main lobe adjustable, almost half of the power contained in the main lobe and the fluctuation of the peak intensity below 15%, which makes them attractive for a variety of applications ranging from micro-machining and imaging to micro-manipulation and laser-assisted guiding.

This work is supported by the National Basic Research Program of China (973 Program) (2014CB340000), the Natural Science Foundations of China (61323001, 61490715, 51403244, and 11304401), and the Natural Science Foundation of Guangdong Province (2014A030313104).